\newcommand\etal{\it et al. \rm}

\documentclass[12pt]{article}
\usepackage{aaspp4}
\usepackage{epsfig}
\begin{document}
\title{Lopsided Galaxies, Weak Interactions and Boosting the Star Formation
Rate }

\author{Gregory Rudnick, Hans-Walter Rix \footnote{Alfred P. Sloan Fellow}$^{,}$\footnote{Current address Max-Planck-Institut f\"{u}r Astronomie, Heidelberg, Germany}, Robert C. Kennicutt, Jr. }

\affil{Steward Observatory, University of Arizona, Tucson AZ 85721}

\begin{abstract}
To investigate the link between weak tidal interactions in disk
galaxies and the boosting of their recent star formation, we obtain
images and spatially integrated spectra ($3615~{\rm \AA} \leq \lambda
\leq 5315~{\rm \AA}$) for 40 late-type spiral galaxies (Sab-Sbc) with
varying degrees of lopsidedness (a dynamical indicator of weak
interactions).  We quantify lopsidedness as the amplitude $\langle
\tilde{A}_1\rangle$, of the $m=1$ Fourier component of the azimuthal
surface brightness distribution, averaged over a range of radii.  The
median spectrum of the most lopsided galaxies shows strong evidence
for a more prominent young stellar population (i.e. strong Balmer
absorption, strong nebular emission, a weak $4000~{\rm \AA}$ break and
a blue continuum) when compared to the median spectrum of the most
symmetric galaxies.  We compare the young stellar content, quantified
by $EW(H\delta_{abs})$ and the strength of the $4000~{\rm \AA}$ break
($D_{4000}$), with lopsidedness and find a $3-4\sigma$ correlation
between the two.  We also find a $3.2\sigma$ correlation between
$EW(H\beta_{emission})$ and lopsidedness.  Using the evolutionary
population synthesis code of Bruzual \& Charlot we model the spectra
as an ``underlying population'' and a superimposed ``boost
population'' with the aim of constraining the fractional boost in the
SFR averaged over the past $0.5$ Gyr (the characteristic lifetime of
lopsidedness).  From the difference in both $EW(H\delta_{abs})$ and
the strength of the $4000~{\rm \AA}$ break ($D_{4000}$) between the
most and least symmetric thirds of our sample, we infer that $\sim 1
\times 10^9 M_{\odot}$ of stars are formed over the duration of a
lopsided event in addition to the ``underlying'' SFH (assuming a final
galactic stellar mass of $10^{10} M_{\odot}$).  This corresponds to a
factor of $ 8$ increase in the SFR over the past $5 \times 10^8$
years.  For the nuclear spectra, all of the above correlations except
$D_{4000}$ vs. $\langle \tilde{A}_1\rangle$ are weaker than for the
disk, indicating that in lopsided galaxies, the SF boost is not
dominated by the nucleus.
\end{abstract}


\section{INTRODUCTION}
Galaxies do not live isolated lives, but exist in the tidal fields of
their environment.  Arp (1966), in his Atlas of Peculiar Galaxies, lay
the observational groundwork for the modern study of interacting
galaxy systems by identifying many "peculiar" systems, later
interpreted as various stages of major galaxy mergers.  Strong
galaxy-galaxy interactions may dramatically alter the stellar
populations (e.g. Larson \& Tinsley 1978; Kennicutt \etal 1987; Turner
1998; Kennicutt 1998), morphology (e.g. Toomre \& Toomre 1972;
Hernquist, Heyl \& Spergel 1993) and kinematics of galaxies
(e.g. Toomre \& Toomre 1972; Barnes \& Hernquist 1992) driving
evolution along the Hubble sequence.  Massive mergers are also capable
of funneling gas into the center of galaxies causing nuclear
starbursts (Barnes \& Hernquist 1991; Mihos, Richstone \& Bothun 1992;
Barnes \& Hernquist 1996) and QSO activity (e.g. Sanders \etal 1988).
At the present epoch, however, major mergers are fairly rare events
(e.g. Kennicutt \etal 1987) and their broad evolutionary importance is
unclear.

Minor mergers and, in general, weak tidal interactions between
 galaxies occur with much higher frequency than major ones (e.g. Lacey
 \& Cole 1993).  By weak interactions we mean those which do not
 destroy the disk of the ``target'' spiral.  Hierarchical structure
 formation models (e.g. cold dark matter) predict that the merging
 histories for high mass objects today contained multiple low mass
 accretion events in their past (e.g. Lacey \& Cole 1993).  The
 specific roles which weak interactions play in the evolution of
 galaxies, however, is uncertain.  Weak interactions may cause disk
 heating (e.g. Toth \& Ostriker 1992; Quinn, Hernquist \& Fullagar
 1993) and satellite remnants may build up the stellar halo
 (e.g. Searle \& Zinn 1978; Johnston, Hernquist \& Bolte 1996).
 Kennicutt \etal (1987) studied the relation between interaction
 strength and star formation by making a comparison between isolated
 galaxies, close pairs, and galaxies from the Arp Atlas.  They found
 that close pairs have larger values of $EW(H\alpha_{em})$,
 i.e. higher star formation rates (SFR) than isolated galaxies.  While
 pair spacing is weakly correlated with the SFR, they could not
 determine the specific role of interaction strength on the SFR.
 Hashimoto \etal (1998) and Allam \etal (1999) both studied the Hubble
 type specific effects of environment on the SFR in galaxies.  They
 found that the SFR/{\it mass of existing stars} was inversely
 proportional to the local galaxy density.  They postulate that the
 anti-correlation is due partly to gas stripping and due partly to the
 anti-correlation of the merger cross-section with the galaxy-galaxy
 velocity dispersion.

There is also evidence that interactions excite nuclear activity.  In
 their close pair and strongly interacting sample Kennicutt {\it et
 al.} (1987) found a strong correlation between $H\alpha$ emission in
 the disk and that in the nucleus.  Such a correlation between disk
 and nuclear emission is supported by theoretical work; Mihos \&
 Hernquist (1994) and Hernquist \& Mihos (1995) demonstrated that minor
 interactions form bar instabilities in the disk which in turn funnel
 large amounts of gas into the nucleus.  The effectiveness of this
 process is suppressed by the presence of a dense bulge, which
 prevents bar formation.  Due to the numerical expense in computing
 high resolution N-body/SPH (collisionless particle/smoothed particle
 hydrodynamics) models, the exact interaction parameters which result
 in such activity are uncertain.

Weak interactions may also manifest themselves as kinematic or
structural irregularities.  Roughly $50\%$ of all spiral galaxies have
asymmetric HI profiles and rotation curves (Baldwin, Lynden-Bell \&
Sancisi 1980; Richter \& Sancisi 1994; Haynes \etal 1998).  Baldwin
\etal (1980) postulated that these asymmetries are caused by weak
interactions in the galaxy's past or by lopsided orbits.  Barton \etal
(1999) examined the optical rotation curves of a set of observed and
simulated interacting disk galaxies.  They showed that interactions
can cause large scale, time dependent asymmetries in the rotation
curves of their sample galaxies.  Swaters \etal (1999) studied the
kinematic asymmetries present in two galaxies lopsided in their
optical and HI distributions.  They qualitatively reproduced the
kinematic asymmetries by placing closed orbits in mildly lopsided
potential.

A dynamical indicator of weak interactions may be ``lopsidedness.''
In the context of this paper (following Rudnick \& Rix 1998; hereafter
RR98), lopsidedness is defined as a bulk asymmetry in the {\it mass}
distribution of a galactic disk.  Surveys for lopsidedness in the
stellar light of galaxies were first carried out by Rix \& Zaritsky
(1995; hereafter RZ95) and Zaritsky \& Rix (1997; hereafter ZR97).
Using near-IR surface photometry of face-on spiral galaxies (spanning
all Hubble types) they examined the magnitude of the $m=1$ azimuthal
Fourier component of the I and K-band surface brightness, thus
characterizing the global asymmetry of the stellar light.  RZ95 and
ZR97 found that a quarter of the galaxies in their sample were
significantly lopsided.  Using a larger, magnitude limited sample
restricted to early type disks (S0 to Sab) and imaged in the R-band,
RR98 found that the fraction of significantly lopsided early type
disks is identical to that for late-type disks.  RR98 convincingly
demonstrated that lopsidedness is not an effect of dust, but is in
fact the asymmetric distribution of the light from old stars and hence
from the {\it stellar mass} in the disk.

Some theoretical work has been done in examining long lived $m=1$
modes (Syer \& Tremaine 1996; Zang \& Hohl 1978; Sellwood \& Merritt
1994), little convincing evidence however has been put forth to show
that isolated galaxies will form stable $m=1$ modes without external
perturbations or significant counter-rotating populations.  Without
invoking the special cases above, long lived lopsidedness is possible
if the disk resides in a lopsided potential.  The question remains
however: how is a lopsided potential created/maintained?  Numerical
simulations of hyperbolic encounters between disk galaxies fail to
produce $m=1$ modes of amplitude $>10\%$ without destroying the
pre-existing stellar disk (Naab, T.; private communication).  Minor
mergers and possibly some weak interactions therefore remain as the
most probable cause of lopsidedness (RR98).  Recent work has shown
that perturbations in the outer halo of a galaxy may be amplified and
even transmitted down into the disk (Weinberg 1994).  Work by Walker,
Mihos \& Hernquist (1996) and ZR97 showed that the type and magnitude
of lopsidedness seen in RZ95, ZR97 and RR98 is comparable to the
result of the accretion of a small satellite, if the mass ratio with
the main galaxy is $\approx 1/10$.  In a preliminary study (i.e. a
rigid halo with no dynamical friction) Levine and Sparke (1998) showed
that lopsided galaxies may be formed by disks orbiting off center and
retrograde in a flat-cored, dark matter dominated halo.  They
postulated that a galaxy may be pushed off center by a satellite
accretion.

Using phase mixing arguments (Baldwin \etal 1980; RZ95) and analysis
 of N-body simulations (Walker \etal 1996; ZR97) the lifetime of
 lopsided features has been estimated at $t_{lop}\approx 1$ Gyr.  That
 lopsidedness is transient ($t_{lop} \ll t_{Hubble}$) yet common,
 requires that it must be recurring and therefore lopsidedness may
 have significant evolutionary consequences.

The current paper focuses on the impact that minor mergers (observed
as lopsidedness) may have on boosting the SFR and the recent star
formation history (SFH) of disk galaxies.  For the purpose of this
discussion, we will assume that lopsidedness is caused by minor
mergers.  Regardless of what causes lopsidedness however, the
perturbation in the gravitational potential manifestly exists and
therefore may affect the gas in the galaxy to such a degree as to
boost the SFR.  Indeed, ZR97 find that lopsidedness is correlated (at
$\geq 96\%$ confidence) with the ``excess'' of blue luminosity (over
what is predicted by the Tully-Fisher relation).  Modeling the
integrated spectral evolution of starbursts using evolutionary
population synthesis (EPS) codes has been been well studied
(e.g. Couch and Sharples 1987; Barger \etal 1996; Turner 1998) and
despite its limitations, is a useful tool in determining the relative
SFH over the past $1$ Gyr.  The same techniques used to probe the SFH
in massive starbursts should also work to probe the recent SFH in the
putative mini-bursts which we seek to study.  By comparing measured
indicators of recent SF (e.g. $EW(H\delta_{abs})$, $4000~{\rm \AA}$
break strength, A star content), to the same indicators derived from
the EPS models, we will place limits on the mini-burst mass and
duration.

We have obtained spatially integrated spectra of a sample of 40 late
 type spiral galaxies (Sab-Sbc) of varying degrees of lopsidedness
 with the intent of using their relative stellar populations (as
 determined from stellar template fitting and EPS models) to determine
 their recent SF histories.  Unlike the mass-normalized blue light
 excess, $\Delta B$ used in ZR97, our method operates independently of
 assumptions about a galaxy's mass, inclination or luminosity.  In
 addition to probing the recent ($\leq 1$ Gyr) SFH with studies of
 the stellar continuum we probe the current SFR by measuring the
 integrated Balmer line emission strengths (e.g. Kennicutt \etal
 1994).

The layout of the paper is as follows.  In \S2 we discuss the sample
 selection, observations, data reduction and determination of galaxy
 lopsidedness; In \S3 we examine our methods for determining the
 current SFR and recent SFH via the measurement of emission and
 stellar continuum properties as a function of lopsidedness.  The
 discussion of the significance of these results, including the
 correlation of the boost parameters with other galaxy characteristics
 and the impact of our results on previous works (i.e. RZ95,ZR97 \&
 RR98) is contained in \S4.  In \S5 we present a summary and possible
 directions for future work.

\section{THE DATA}
\subsection{Sample Selection}
To build a sample of galaxies with varying degrees of lopsidedness, we
imaged a large number of galaxies ($N_{gal} \geq 100$) taken from the
RC3 catalog (De Vaucouleurs et al.  1991), selected according to the
following criteria: apparent blue magnitude m$_B\leq$14, redshift
cz$\leq$10,000 km/s, axis ratio $b/a \geq$0.64 ($50^\circ \leq i
\leq0^\circ$), de Vaucouleurs type $ab \rightarrow bc$, and a maximum
diameter of 4$\arcmin$.  The median diameter of the galaxies in our
sample was 1.8$\arcmin$.  The magnitude and redshift limits were
chosen to minimize the required exposure times.  The axis ratio of the
galaxies was constrained because it is hard to measure an azimuthal
$m=1$ component in a highly inclined galaxy.  Once imaging was
obtained and lopsidedness determined for each galaxy (see \S{\it
2.2.1}), we constructed a sample for spectroscopy consisting of 40 of
our imaged galaxies (see Table 1).  These were selected to give the
sample equal numbers of lopsided and symmetric targets.

Our sample is partially selected according to Hubble type, and we must
 explore the effects which morphological evolution induced by
 lopsidedness may have on our conclusions.  Walker \etal (1996)
 suggested that minor mergers increase bulge size, heat the galactic
 disk vertically and consume a large fraction of the galaxy's gas,
 eventually resulting in a low post-merger SFR. These two effects may
 drive galaxies towards earlier Hubble type after they experience
 minor mergers.  Evolutionary processes such as these however become
 dominant either after the expected lifetime of lopsidedness ($t\geq
 1$ Gyr), or after repeated merger events (Walker \etal 1996).  During
 the interaction itself the irregularity in structure made manifest by
 lopsidedness as well as the creation of spiral arms via tidal
 perturbations will temporarily move a galaxy later in Hubble type.
 This will effectively push lopsided early type disk galaxies into our
 sample while pushing those of later type out of it.  Due to their
 lower gas masses (Roberts \& Haynes 1994), early type spirals have
 less potential for a large absolute increase in their SFR than late
 types.  Small boosts, however, may be easily noticeable against the
 typically older stellar population of early type disks.  The exact
 interplay of these two effects may bias our measurement of the
 relation between SFR and lopsidedness.

\subsection{Observations}
\subsubsection{Imaging}
Our imaging data were obtained during runs at Steward Observatory's
 2.3-m Bok reflector on Kitt Peak (1997 November 6-7 and 1998 February
 1-2) and at its 61-inch (1.5-m) reflector on Mt. Bigelow (1998 May
 15-18).  The CCD pixel scales at the Bok reflector and Bigelow
 reflector were $0\farcs4{\rm pixel}^{-1}$ with fields of view
 $6\farcm8 \times 6\farcm8$ and $3\farcm4 \times 3\farcm4$
 respectively.  The median seeing at the 2.3-m for the November and
 February runs were 1\farcs3 and 1\farcs5, respectively, while the
 median seeing at the 1.5-m in May was 1\farcs3.

A Nearly-Mould $R-$band filter ($\lambda_{center}=650nm$) was used at
 the 2.3-m and a Kron-Cousins R-band filter with
 $\lambda_{center}=650nm$ was used at the 1.5-m. The effects of
 wavelength on observed lopsidedness are discussed in \S2.1 of RR98,
 and found not to be critical.  

To determine the lopsidedness of a galaxy we perform an azimuthal
Fourier decomposition of the R-band surface brightness, as in RR98:
\begin{equation} I(R_m,\phi) = a_o \{ 1 + \sum_{j=1}^N a_j
e^{-i[j(\phi_j-\phi_j^o)] } \},\end{equation} where for each radius,
$\left|a_o\right|(R)$ is the average intensity and
$\left|a_1\right|(R)$ describes the lopsidedness.  We define the
luminosity normalized quantities $A_1~\equiv~a_1/a_0$.  Instead of
$A_1(R)$, we use $\tilde{A}_1(R)$ (the error corrected value which
accounts for the positive definite nature of our measurements and the
presence of errors; see RR98 for details) as our measure of asymmetry.
We calculate the mean asymmetry of each galaxy, $\langle
\tilde{A}_1\rangle$ (see Table 2), from 1.5 to 2.5 disk scale lengths
using the weighted average described in RR98.

\subsubsection{Spectroscopy}
Spectra were obtained with the Bollers \& Chivens Spectrograph at the
2.3-m Bok reflector during the nights, 1998 March 22-25, 1998 May 25-28,
and 1998 June 29.  We used a $400~gmm^{-1}$ in $2^{nd}$ order, blazed
at $3753~{\rm \AA}$, and a $2.5\arcsec$ slit, resulting in a
resolution of~$\approx1500$ and a wavelength range of $3600~{\rm \AA}
\lesssim \lambda \lesssim 5300~{\rm \AA}$.  This range includes the
entire Balmer series redward to $H\beta$, the $4000~{\rm \AA}$ break,
Ca H+K doublet, [\ion{O}{2}]$\lambda\lambda 3726,3729~{\rm \AA}$ and
[\ion{O}{3}]$\lambda\lambda 4959,5007~{\rm \AA}$.  To reduce read
noise, the CCD was binned in the spatial direction; on 22 March we
binned by 2 for a resultant pixel scale of $1.67\arcsec{\rm
pixel}^{-1}$ while on all other nights we binned by 4 for a pixel
scale of $3.33\arcsec{\rm~pixel}^{-1}$.  A CuSO$_4$ filter was used
to block $1^{st}$ order light.  Aside from using standard stars to
calibrate the relative spectral response of the instrument, no
absolute flux calibration was attempted. This was done partly because
of the non-photometric conditions of some of our nights and partly due
to the independence of our analysis methods on absolute flux levels.

Following Kennicutt (1992), we obtained spatially integrated spectra of the
 galaxies by repeatedly scanning the slit across the galaxy between
 $-2.5R_{exp} \leq x \leq 2.5R_{exp}$.  For each galaxy we obtained 2
 exposures of 25 minutes each.  To isolate the disk contributions to
 the integrated spectra, we also obtained a 5 minute exposure of the
 nucleus for each galaxy for later subtraction.

\subsection{Reduction}
\subsubsection{Images}
The basic image reduction was carried out with standard
 IRAF\footnote{IRAF
 is distributed by the National Optical Astronomical Observatories,
 which are operated by AURA, Inc. under contract to the NSF.}
 routines.  The total dark current was $\leq 1 {\rm
 e^-/exposure/pixel}$ and so it was ignored.  The images were
 flat-fielded with a combination of twilight and smoothed night sky
 flats.  High $S/N$ twilight flats ($N\approx 5$) were used to take
 out small-scale variations and the lower $S/N$ smoothed night sky
 flats ($N\approx 6$) were used to take out large-scale sensitivity
 fluctuations.  The large-scale flat-field quality was determined by
 measuring variance in the median sky level at the four corners of the
 images.  In all images where the galaxy was small in the field of
 view, the images were found to be flattened to better than $1\%$.  In
 some cases, the large size of the galaxies in comparison to the field
 of view at the 1.5-m telescope precluded such an estimate.
 
Point sources in the images were selected using DAOFIND, and
 surrounding pixels were excised in the subsequent analysis to a
 radius where the stellar point spread function declined to the level
 of the sky.

\subsubsection{Spectra}
The basic spectral reduction was carried out with IRAF routines.  The
 measured dark current from multiple 25 minute dark exposures was
 $\leq 2 {\rm e^-/exp/pixel}$ and so was not accounted for.
 Small-scale variations were removed with a combined series of 2
 min. dome flats.  Twilight flats were fit in the spatial direction
 with a $5-7^{th}$ order cubic spline at several different positions
 in the spectral direction to construct a map of the slit response as
 a function of wavelength.  Non-linear pixels were interpolated over
 using a bad pixel mask generated from the ratio between a combined
 series of 2 minute and 10 second dome flats.

Removing cosmic rays from spectra without accidently removing emission
 lines is best automated by using the shape of the spectrum itself in
 the cleaning process: for a given column (spatial direction), we
 medianed together the target column with the eight columns on either
 side to construct a slit profile, $I_{slit}$.  Using $\chi^2$
 minimization, we fit the target column with the the following
 function: \begin{equation} I_{model}(y)=a_1 I_{slit}(y) + a_2 y +
 a_3.\end{equation} where the $a_3$ term accounts for the sky
 background.  Pixels deviating by more than 7 $\sigma$ from the best
 fit are replaced with the value of the best fit model at that point.
 Flagged segments larger than the spectral resolution of the
 instrument were not removed so as to avoid the accidental cleaning of
 emission lines. This algorithm is very efficient and, once a suitable
 set of parameters (i.e. $N_{med}$, threshold) has been chosen, can
 remove almost all of the cosmic rays on the chip.

After cosmic ray removal, the spectra were rectified using He-Ar lamps
 taken at various times during the night.  Background contributions
 were then subtracted.  To extract the spectra, we summed the number
 of center rows which corresponded to $\pm 2.5R_{exp}$ for the disk
 spectra and extracted the center row only for the nuclear spectra.
 Using standard stars taken during the night, we then removed the
 spectral response of the system from the 1D, extracted spectra.
 Finally, the wavelength scale of each spectra was shifted to the rest
 frame of the galaxy.

The guided nuclear exposures were scaled by the effective exposure
 times on the nucleus during the drift scanning.  These scaled
 exposures were then subtracted from the drift spectra.  In this way,
 we separated the disk and nuclear contributions.  Finally, all
 spectra were scaled to their median flux level.

\subsection{Errors}
The error calculation for $\langle \tilde{A}_1\rangle$ is described in
 RR98; it accounts for both photon statistics and systematic
 flat-fielding uncertainties.

Using Poisson statistics to describe the error in a raw spectrum we
 created an ``error spectrum'' and performed on it all of the standard
 reduction steps discussed in \S{2.3.2} except for the cosmic ray
 cleaning.  After normalizing the ``science spectra,'' the ``error
 spectra'' were scaled by the median of the ``science spectra'' to
 form a detailed representation of the error at each pixel.  The
 ``error spectra'' were used in all of the following analysis steps.

\section{SPECTRAL ANALYSIS \& STAR FORMATION HISTORY DIAGNOSTICS}
By examining spatially integrated spectral characteristics as a
function of $\langle \tilde{A}_1\rangle$, we can determine whether
lopsidedness affects the SF histories of galaxies.  It is difficult,
even in major mergers, to invert optical spectrophotometry into a SFH
estimate (e.g. Turner 1998).  However, the relative strength of the
current and recent SFR in different galaxies may be studied with
moderate $S/N$, non flux-calibrated spectra (e.g. Couch \& Sharples
1987; Barger {\it et al.} 1996).  The equivalent width of $H\alpha$ in
emission ($EW(H\alpha_{em})$) integrated over the whole disk is a
robust measure of the current global SFR in terms of the previously
formed stars (Kennicutt et al. 1994).  Much of the Balmer emission inn
a galactic disk comes from the HII regions seen in SFR regions.  In
this regime, the Balmer decrement relates the emission flux in
$H\beta$ to that in $H\alpha$.  If the continuum level at these two
wavelengths is similar, then the decrement will also relate the $EWs$
of the two lines.  Therefore, the absorption corrected
$EW(H\beta_{em})$ (see \S3.2) serves as an indirect
indicator of the current SFR in the disk.

The likely lifetime of lopsidedness has been estimated as
$t_{lop}\approx 1$ Gyr (Baldwin \etal 1980; RZ95; ZR97).  To examine
SF on these timescales, we need a SF tracer with a comparable
lifetime.  Main sequence A-stars have lifetimes of $\approx 0.5$ Gyr
(Clayton 1983), have strong spectral signatures (e.g. strong Balmer
lines, a blue continuum, and a weak $4000~{\rm \AA}$ break), and so
serve as appropriate probes of the recent SFH.

\subsection{Fitting Stellar Templates}
Fitting population synthesis models (e.g. Bruzual \& Charlot 1993) to
our spectra can give us a measure of the recent SFH.  However for four
reasons we choose to simply fit with empirical stellar templates
(Jacoby \etal 1984): 1) The spectral resolution of available
population synthesis models ($\approx 10~{\rm \AA}$) is significantly
less than that of our spectra ($3~{\rm \AA}$). 2) The empirical
templates of Jacoby \etal (1984) have a spectral resolution ($4.5~{\rm \AA}$)
slightly less than that of our spectra. 3) An adequate fit to the
spectra can be obtained with a small number of high S/N stellar
templates.  To estimate the recent, relative SFH, we therefore need
not deal with the complexities of the IMF and metallicity of the
stellar populations.

As a qualitative measure of the relative contributions to our spectra
from young and old stars, we synthesize the global absorption spectra
of the galaxies in our sample with a linear combination of two stellar
templates, an A0V and a G0III spectra from the Jacoby \etal (1984)
stellar library.  We also individually fit and subtract a $3^{rd}$
order polynomial of zero mean from the spectra and from each stellar
template to correct for color terms (e.g. from calibration errors,
from approximating the spectra with only two stellar templates).  The
spectra (with the polynomial subtracted) are fit by $\chi^2$
minimization with the following model:\begin{equation} I_{model}
(\lambda)=(C_{A0V}I_{A0V}(\lambda)) \otimes G(\sigma) + (C_{G0III}
I_{G0III}(\lambda)) \otimes G(\sigma) - I_{poly},
\end{equation} where $G(\sigma)$ and the weights, C were determined
iteratively.  $C_{A0V}$ and $C_{G0III}$ are the weights for the
normalized stellar template spectra, $I_{A0V}$ and $I_{G0III}$.
$I_{poly}$ is the sum of the weighted template polynomial components,
$G(\sigma)$ represents the Doppler broadening of the stars which is
convolved with the stellar templates, and $\otimes$ is the convolution
operator.  For more details on the fitting procedure see Rix \etal
(1995) and Turner (1998).

Independent of continuum slope, there is a unique set of line shapes
and strengths for stars of each spectral type.  The polynomial fit
minimizes the effects of a global continuum slope on our best fit
solution so that we are instead performing a global fit to the
spectral features ($4000~{\rm \AA}$ break, Ca H$+$K, Balmer series,
etc.)  We find that with only two, polynomial subtracted, stellar
templates, we are able to consistently achieve fits
with $\chi^2_{\nu} \leq 2$.

\subsection{Emission Lines}
Our template fit to the galaxy spectra should reflect only the stellar
populations in the galaxy, not interstellar emission. Therefore, we
first fit all portions of spectra, omitting expected emission regions,
and use the residual of the best model fit to construct an emission
spectrum.  We remove the large scale variations in the residual by
fitting it with a high order polynomial ($\approx 50$) at all
locations where no emission is expected.  We then fit each emission
feature with a Gaussian.  To isolate the stellar continuum, we
subtract these gaussian components from the original spectra.  We then
re-fit our $I_{model}(\lambda)$ to this cleaned, pure absorption
spectrum to determine $I_{model}^{best} (\lambda)$.  This process is
illustrated in Figure 1.

We measure $EW(H\beta_{em})$ (see Table 2) from the complementary
absorption corrected emission line spectra for all of our galaxies.  A
linear continuum was fit on either side of $H\beta$ ($4720~{\rm \AA}
\leq \lambda_B \leq 4800~{\rm \AA};4900~{\rm \AA} \leq \lambda_R \leq
4940$).  We then measured the equivalent width of the emission line
integrating from $4843~{\rm \AA} \leq \lambda_{line} \leq 4883~{\rm
\AA}$.  The boundaries of our integration were chosen by visual
inspection to minimize noise contributions from the continuum while
maximizing the amount of line flux.

\subsection{Quantifying the A-star Fraction}
We can quantify the relative A star abundances by simply using the
value of $C_{A0V}$ (see Table 2) in $I_{model}^{best} (\lambda)$.
Because the basis templates are normalized to their median fluxes (as
are the data,) the scaling factor of the individual templates gives a
measure of how much A-stars contribute to the integrated spectra.

Balmer absorption lines are strongest in A-stars, the $4000~{\rm \AA}$
break is weak, and so we may also use these two features to measure
the recent SFH (Couch \& Sharples 1987; Barger \etal 1996).  We use
the equivalent width of $H\delta$ in absorption ($EW(H\delta_{abs})$)
as our indicator of Balmer line strength in order to minimize emission
contamination and sample a relatively isolated region of the spectrum.
A linear continuum was fit on either side of $H\delta$ ($4000~{\rm
\AA} \leq \lambda_B \leq 4050~{\rm \AA} ; 4150~{\rm \AA} \leq
\lambda_R \leq 4250~{\rm \AA}$).  We measure the $EW$ itself from
$4080~{\rm \AA} \leq \lambda_{line} \leq 4120~{\rm \AA}$.

Instead of measuring $EW(H\delta_{abs})$ directly from the spectrum,
we decide to measure $EW_{mod}(H\delta_{abs})$ (see Table 2) of the
best matching template spectrum, $I_{model}^{best} (\lambda)$.  These
template spectra are in general a good match to the data and provide
an essentially noiseless estimate of $EW(H\delta_{abs})$ with a value
derived from the best fit to a broad wavelength range.  When measuring
$EW(H\delta_{abs})$ directly from the spectra, the relation between
$EW(H\delta_{abs})$ and $\langle \tilde{A}_1\rangle $ (see \S4.3)
clearly remains, but the scatter in $EW(H\delta_{abs})$ at a given
$\langle \tilde{A}_1\rangle $ is slightly larger.

To further parameterize the recent SFH, we use the strength of the
$4000~{\rm \AA}$ break ($D_{4000}$) (see Table 2).  Our measure of the
break strength is defined as: \begin{equation}
D_{4000}=\frac{\int^{4250}_{4050} f_{\lambda}
d\lambda}{\int^{3950}_{3750} f_{\lambda} d\lambda}.
\end{equation} The suppression of the continuum blueward of $4000~{\rm
\AA}$ manifested in the break is caused by the combined absorption
from the Ca H$+$K and $H\epsilon$ absorption line.  As young, massive
stars (with blue continua and weak metal lines) contribute more to the
overall spectrum, $D_{4000}$ decreases.

\section{RESULTS AND DISCUSSION}
\subsection{Overall Spectral Characteristics}
To examine qualitatively the link between lopsidedness and SFR, we
 constructed composite disk spectra from the median of our nine most
 symmetric and nine most lopsided galaxies, and these we show in
 Figure 2.  The lopsided composite spectrum has a much bluer
 continuum, stronger Balmer lines, stronger emission and a shallower
 $4000~{\rm \AA}$ break than its symmetric counterpart regardless
 of exactly how many galaxies we include in the median spectrum.
 This result indicates that lopsidedness is in fact correlated with
 the recent SFH.  However, a quantitative treatment of the relation is
 still needed.

\subsection{Is the Current Star Formation Correlated with Lopsidedness?}
Figure 3 shows the relation between $EW(H\beta_{em})$ and $\langle
 \tilde{A}_1\rangle $ for the entire spectroscopic sample.  Using a
 Spearman-rank test, we find that the distribution of
 $EW(H\beta_{em})$ vs. $\langle \tilde{A}_1\rangle $ deviates from a
 random one at the $99.9\%$ level.  The median $EW(H\beta_{em})$ of
 our nine most symmetric galaxies is $-3.2~{\rm \AA}$ while the median
 for our nine most lopsided galaxies is $-6.3~{\rm \AA}$.  The
 correlation of $EW(H\beta_{em})$, and hence the current SFR, with
 $\langle \tilde{A}_1\rangle$ is clearly shown in Fig. 3.  This
 implies that an elevation in the instantaneous SFR occurs over the
 same timescales as lopsidedness, lasting maybe as long as $1$ Gyr
 after the onset of lopsidedness.

\subsection{Is Recent Star Formation Correlated with Lopsidedness?}
 Figure 4 shows the relation of A-star content,
 $EW_{mod}(H\delta_{abs})$, and $D_{4000}$ with $\langle \tilde{A}_1
 \rangle$.  They deviate from random distributions at the $99.997\%$,
 $99.99\%$, and $99.68\%$ levels respectively.  Our sample is not
 large enough to study these correlations separately for different
 Hubble types; instead we analyze the change in spectral properties of
 the sample as a whole.  Figure 5 shows that the data are
 differentiated by lopsidedness in the $EW_{mod}(H\delta_{abs})$
 vs. $D_{4000}$ plane.  The large scatter is not surprising, as even
 within a single Hubble type there is a considerable variation in the
 specific SFHs (Kennicutt \etal 1994).  Between the most lopsided and
 symmetric $1/3$ of our galaxies the median values of
 $EW_{mod}(H\delta_{abs})$ and $D_{4000}$ differ by, $\Delta
 EW_{mod}(H\delta_{abs})=2.1\pm1.0~{\rm \AA}$ and $\Delta
 D_{4000}=0.24\pm0.01$, respectively.  This defines a boost vector
 (Fig. 5) whose direction and magnitude characterize the difference in
 spectral indices between a symmetric and lopsided galaxy in our
 sample.

\subsection{Constraining the Boost Mass}
A number of efforts have shown that the detailed SFH, even over the
last $10^9$ years, cannot be determined unambiguously from integrated
spectra (e.g. Turner 1998 and references therein; Leonardi \& Rose
1996).  Instead, we restrict ourselves to a set of model SFHs
consisting of a ``normal'' or ``underlying'' spiral galaxy SFH, which
in lopsided galaxies has been boosted in the past by a factor
$C_{boost}$.  We then attempt to characterize the boost strength
,$C_{boost}$, and an associated timescale.  Using the EPS code of
Bruzual \& Charlot, (GISSEL 1995; in preparation) we construct SEDs
representing the light from the underlying stellar populations of
galaxies in our sample.  We specify the SFH by the birthrate parameter
$b$ (Scalo 1986), where
\begin{equation} b=\frac{{\rm SFR}_{current}}{\langle {\rm SFR}_{past}
\rangle}.\end{equation} Assuming an exponential SFH, i.e. $SFR(t)
\propto \exp \left[ -t/\tau(b) \right ]$, each $b$ has an associated
timescale, $\tau(b)$ defined through \begin{equation}\exp \left[
\frac{t_{gal}}{\tau(b)}\right ] -1 = \frac{t_{gal}}{\tau(b)}
\frac{1}{b},\end{equation} with $t_{gal}$ as the age of the galaxy.
We adopt an age of $t_{gal}\approx 10$ Gyr and a metallicity of
$Z=Z_{\odot}$.  We also use the Salpeter (1955) IMF with a mass range
of $0.1M_{\sun}-125M_{\sun}$.  This IMF provides a better fit than
does the Scalo IMF (1986) to the global photoionization rates and to
the colors of galactic disks (Kennicutt et al. 1994).  Applying the
Scalo IMF results in slightly lower fractional boost masses than the
Salpeter IMF, but the IMF choice does not qualitatively affect our
conclusions.  The model SED is then given by the integral:
\begin{equation} F \Bigl ( \tau(b),\lambda \Bigr ) =\int^{t_{gal}}_0 S(t',\lambda)
r\Bigl ( \tau(b),t' \Bigr ) dt', \end{equation} where \begin{equation}
r \Bigl ( \tau(b),t \Bigr ) =\beta(b) \exp \left[ -\frac{t}{\tau(b)}
\right], \end{equation} where $\beta(b)$ is the Hubble type dependent
initial SFR of the galaxy used to normalize the present day masses to
the same value, and where $S(t',\lambda)$ is the SED for a population
$t'$ years after a delta function burst (Bruzual \& Charlot's simple
starburst model).

We account for the change in the spectral properties by {\it adding}
an exponentially declining increase ,or ``boost'' in SF on top of the
underlying galactic SFH.  We assume that the boost originated about
$5\times 10^8$ years ago, comparable to the presumed dynamical age of
the boost and the A-stars seen in Figure 2.  The model SED for a
galaxy with a boost population is given by: \begin{equation} F\Bigl (
\tau(b),\lambda \Bigr ) =\int^{t_{gal}}_0 S(t',\lambda) \beta(b) \exp
\left[ -\frac{t'}{\tau(b)} \right] dt' +
\int^{t_{boost}}_{0} S(t',\lambda) C_{boost} \exp \left[
-\frac{t'}{\tau_{boost}} \right] dt'
\end{equation}.

For each Hubble type, i.e. for each $\tau(b)$ we construct model
 tracks by varying $C_{boost}$ and $\tau_{boost}$ in Equation (9),
 effectively adding differing boosts to the unperturbed galaxy SFH.
 As shown in Fig. 6, moving the underlying population later in Hubble
 type, or to a larger $\tau(b)$, shifts the tracks to greater values
 of $EW_{mod}(H\delta_{abs})$ and smaller values of $D_{4000}$.  For a
 given Hubble type, varying the boost timescale changes the slope of
 the track of increasing boost mass.

It is interesting to note that our empirical boost vector cannot be
matched directly with any of the models in the
$EW_{mod}(H\delta_{abs})$ vs. $D_{4000}$ plane, suggesting that the
EPS models may have systematic errors.  Assuming, however, that the
correct models will differ from these by a simple shift in the
$EW_{mod}(H\delta_{abs})$ vs.  $D_{4000}$ plane, we choose the
underlying population model which minimizes the distance (in the
$EW_{mod}(H\delta_{abs})$ vs. $D_{4000}$ plane) to the beginning of
our observed boost vector.  We then match the slope of the boost
vector by varying the boost timescale.  The best match we find was for
$b=0.33$ (Hubble type Sb; Kennicutt et al.  1994) and
$\tau_{boost}=500$ Myr.  For reference, we also show the tracks for
underlying galaxy spectra with $b$ parameters corresponding to Hubble
types of Sab and Sbc ($b=0.17$ and $0.84$ respectively; Kennicutt et
al.  1994).  For each of these Hubble types, we plot tracks
corresponding to boosts with $\tau_{boost}=100$ Myr and
$\tau_{boost}=500$ Myr.  We also plot the $\tau_{boost}=100$ Myr boost
for the $b=0.33$ model.  Measuring the projected length of the boost
vector on the best model, and assuming the galaxy disk has a final
stellar mass of $10^{10}M_{\odot}$, we find that $\sim 1
\times 10^9 M_{\odot}$ of stars have been formed over the the past
$0.5$ Gyr in addition to the ``underlying'' SFH.  This corresponds to
an factor of $\sim 8$ increase in the SFR over the past $5 \times
10^8$ years.

Extinction will have two primary effects on our data.  It will redden
the continuum from the galaxy.  By allowing for a polynomial element
to our template fits, we minimize the effect of the continuum slope on
our results.  Extinction may also affect regions of current SF;
obscuring both emission lines and the continuum from massive young
stars.  We do not make a further account of reddening but acknowledge
that our boost masses may increase if reddening is included.

\subsection{Correlations in Disguise} 
It is important to verify that our result is not an artifact of
correlations between the underlying SFH and lopsidedness.  To
determine whether mass, gas content or Hubble type effects mimic our
relation between lopsidedness and SFR, we examined the distributions
of $EW(H\beta_{em})$, $EW_{mod}(H\delta_{abs})$, $D_{4000}$, A star
content and $\langle \tilde{A}_1\rangle$ vs. $L_B$, $M_{HI}/L_B$ and
T-type (De Vaucouleurs et al. 1991).  A hidden correlation exists if
any of the latter three parameters are correlated both with a SF
indicator and with $\langle \tilde{A}_1\rangle$.

None of the above distributions, except for $EW(H\beta_{em})$ and
 $D_{4000}$ vs. T-type and $M_{HI}/L_B$, have more than a $1.7\sigma$
 probability of being correlated, with a $\leq 2.6\sigma$ significance
 for the four remaining correlations.  Given the known correspondence
 between SFR and Hubble type, mean stellar population and Hubble type,
 and SFR and gas surface density (Kennicutt et al. 1994), these
 correlations with $EW(H\beta_{em})$ and $D_{4000}$ are expected.
 However, none of our canonical galaxy parameters show a correlation
 with lopsidedness at above the $1.2\sigma$ level and we can therefore
 be confident that we are seeing a true correlation between our SF
 indicators and $\langle \tilde{A}_1 \rangle $.

\subsection{Nuclear Properties and Contributions}
As an initial diagnostic of the nuclear contributions, we examine the
 fraction of galactic light originating in the nucleus.  Even without
 absolute spectrophotometry, we can still examine what percentage of
 the total galaxy light is made up of light from the nucleus as long
 as the spectra are calibrated with respect to each other.

The fractional luminosity of the nucleus ranged from $0.1-6.4\%$ with
a median value of $1.7\%$.  The fractional luminosity of the nucleus
is uncorrelated with $\langle \tilde{A}_1 \rangle$.  A Spearman-Rank
test showed that there was a $52\%$ probability that the data was
drawn from a random sample.  Figure 7 shows that while the nuclear
$EW(H\beta_{em})$ vs. $\langle \tilde{A}_1 \rangle$, and
$EW_{mod}(H\delta_{abs})$ vs. $\langle \tilde{A}_1 \rangle$ are
correlated, the relations show more scatter than in the disk.  The
Spearman-Rank test shows that these two relations are less
significantly correlated in the nucleus than in the disk by
$0.4\sigma$ and $0.7\sigma$, respectively.  Only for $D_{4000}$
vs. $\langle \tilde{A}_1 \rangle$ was the correlation greater than in
the disk.

Our data indicate that the enhanced SF in lopsided galaxies occurs in
the disk with equal or greater strength than in the nucleus. This
result appears to run contrary to what numerical simulations suggest
about the reactions of gas in the disk to tidal perturbations from
external sources.  Mihos \& Hernquist (1994) and Hernquist \& Mihos
(1995) show that even high mass ratio mergers ($M_{gal}/M_{sat} > 10$)
can form a bar which torques gas in the disk and funnels much of it
into the nucleus.  However the presence of a dense bulge easily
inhibits bar formation and hence gas is not funneled as effectively.
Another explanation is that star formation is occurring in large
amounts at the nucleus, but is partly masked by extinction.  High
extinction is often found to be coincident with high star formation
regions in mergers and is easily capable of hiding the presence of
young populations (Turner 1998).
 
\subsection{The Frequency of Lopsidedness-- Revisited}
To estimate the volume density of lopsided galaxies from a magnitude
limited sample requires the understanding of any systematic
differences in luminosity between lopsided and symmetric galaxies
(ZR97).  Our best fit boost brightens the galaxies by 1.3, 1.0 and 0.8
magnitudes (in B, V, and R, respectively) over the brightness of the
underlying population.  Lopsided galaxies will hence be
overrepresented in our B-band selected sample, and their volume
density will be $\sim25\%$ of that derived from any magnitude limited
survey.  To compare the frequency of lopsidedness for different Hubble
types (e.g. RR98), we need to understand how this brightening depends
on the ``underlying'' SFH.  Unfortunately, we have too few galaxies in
our sample to discuss Hubble type dependent effects on the boost
strength and luminosity evolution.

At present, the main constraint on the frequency of lopsidedness is
from its frequency in imaging samples ($20\%$; RR98).  Coupled with
our brightening estimates, we find that $\sim 5\%$ of the galaxies in
a volume limited sample will be lopsided.  Lopsidedness is likely
caused by minor mergers, the frequency of which increases with look
back time.  Given that lopsidedness lasts for $\sim 1$ Gyr, we can
estimate that the average galaxy has been lopsided at least once in
its lifetime.

\section{SUMMARY \& CONCLUSION}
To quantify the correlation between the recent SF histories of
present-day spiral galaxies and their global asymmetry, we compare the
integrated spectral properties of late-type spirals of varying
lopsidedness.  We find that the recent ($\leq 0.5$ Gyr) SFH and
current ($\leq 10^7$ years) SFR are both strongly correlated with
$\langle \tilde{A}_1 \rangle$ although there is appreciable scatter in
the individual galaxy-to-galaxy properties.  For $EW(H\beta_{em})$,
reflecting the current SFR, we find a $3.2\sigma$ Spearman-rank
correlation with $\langle \tilde{A}_1 \rangle$.  We fit a combination
of A0V and G0III stellar spectra to our galaxy spectra to quantify the
relative abundance of A-stars in the disk (which traces the SFR within
0.5 Gyrs).  From these best fit model spectra, $I_{model}^{best} (\lambda)$,
we measure a number of spectral indices, and find that
$EW_{mod}(H\delta_{abs})$, $D_{4000}$, and $C_{A0V}$ are correlated
with $\langle \tilde{A}_1 \rangle$ at the $3.9\sigma$, $3.0\sigma$,
and $4.2\sigma$ levels, respectively.

We measure the same spectral indices in the nucleus, and find them
less correlated with $\langle \tilde{A}_1 \rangle$ (except
$D_{4000}$).  Unless a nuclear starburst is obscured, the disk and not
the nucleus is the primary site of the SF increase we see in lopsided
galaxies.  This is in contrast to numerical simulations where minor
mergers funnel gas into the nucleus of galaxies, causing intense
starbursts (Mihos \& Hernquist 1994; Hernquist \& Mihos 1995).  Only
by the presence of a dense bulge can the formation of a bar, and the
subsequent funneling of gas, be prevented.

To quantify the mass of additional stars formed in lopsided galaxies,
we defined a boost vector in $EW_{mod}(H\delta_{abs})$ vs. $D_{4000}$
space, by comparing the median values of these properties for the most
symmetric third and the most lopsided third of our sample.  We find
$\Delta EW_{mod}(H\delta_{abs}) = 2.1\pm 1.0~{\rm
\AA}$ and $\Delta D_{4000} = 0.024\pm 0.01$.  We fit this vector with
an ``underlying population $+$ boost'' EPS model corresponding to a
progenitor galaxy with $b=0.33$, $\tau_{boost}=500$ Myr, and boost age
of $0.5$ Gyr.  Using this best fit EPS model, we find that $\sim 1
\times 10^9 M_{\odot}$ is formed in the boost in addition to the
``underlying'' SFH (assuming a final stellar mass of $10^{10}
M_{\odot}$).  This is a considerable fraction ($\sim 10\%$) of the
final stellar mass of the galaxy and corresponds to a factor of 8
increase in the SFR over the past $5 \times 10^8$ years.  Given the
increasing merger rates and increasing gas fractions towards higher
redshifts, minor merger induced SF boosts of short duration played an
important role in assembling the present day stellar content of
galaxies.

Finally, we address by how much the frequency of lopsidedness from a
magnitude limited sample is increased by the corresponding luminosity
boost.  Our best fit EPS boost model corresponds to a $\sim 1$
magnitude brightening when galaxies becomes lopsided, increasing their
presence four-fold in magnitude limited samples.  We lack the
statistics however, to examine any Hubble type dependent differences
in the luminosity boost.

It is obvious that more work needs to be done to fully understand the
 cause of lopsidedness as well as the SFH of lopsided galaxies.  To
 quantify the Hubble type specific boost in the recent SFH, a large
 sample should be obtained with significant numbers of galaxies in
 each Hubble type bin.  Since imaging and spectroscopy will be needed
 for this project, a volume limited sample may be constructed which
 bypasses many of the problems encountered when selecting galaxies
 according to an apparent magnitude limit.  Companion searches to
 sufficiently faint magnitudes will help to study the possible link
 between environment and lopsidedness (as caused by weak tidal
 interactions).  With the recent commissioning of large area imaging
 and spectroscopy surveys such as Sloan Digital Sky Survey,
 constructing such a sample will become relatively straightforward.
 
Numerical simulations have shown to be a useful tool in studying the
 evolution of the stellar and gas distributions in minor mergers.
 High resolution simulations with a live halo are crucial for studying
 the detailed response of the disk to the merger (Walker \etal 1996).
 A thorough exploration of interaction parameter space is needed to
 quantify the structural and kinematic response in the stellar and gas
 components.  High resolution N-body studies are also needed to
 explore the global stability of isolated galactic disks.

\acknowledgments
This work was completed with partial support from NSF grants
 AST9870151, AST9421145 and AST9900789.  Greg Rudnick and H.-W. Rix
 would like to thank Nelson Caldwell for many valuable discussions on
 measuring SFHs from integrated spectra.  Greg Rudnick would like to
 thank Craig Kulesa and Christopher Gottbrath for many useful
 discussions in the early hours of the morning.  Greg Rudnick would
 also like to thank Megan Sosey and Chris Gottbrath for assisting with
 our observing program.  Finally, we would like to thank the Steward
 Observatory 2.3-m telescope Operators for their assistance in the
 completion of this project.

{}

\clearpage

\begin{figure}
\epsfig{file=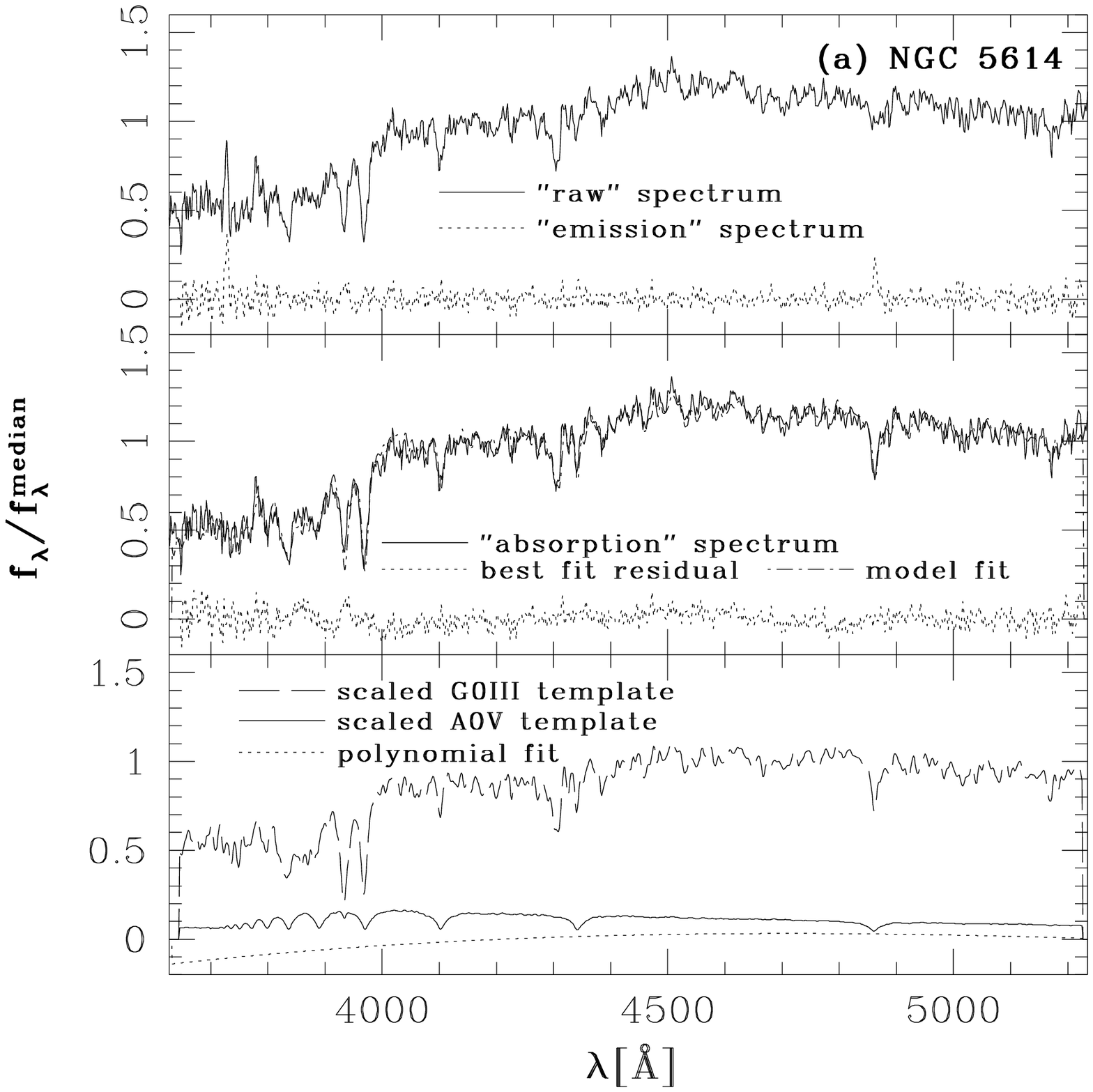,height=6.5in,width=6.5in}
\caption{ 
Empirical model fits (G0III $+$ A0V) for two sample galaxies: (a) NGC
5614, in which the older population (modeled by a G0III star) clearly
dominates the spectrum and (b) NGC 4595, whose stellar light is made
up in large part, of light from young massive stars (modeled by a A0V
star).  The top panel shows the ``disk'' spectrum prior to emission
line decomposition.  It also shows the ``emission'' spectrum created
from the best fit residual of the first pass through the fitting
routine.  The middle panel shows the ``absorption'' spectrum, the best
fit model, $I_{model}^{best} (\lambda)$, and the resulting fit
residual.  The bottom panel gives the weighted, doppler broadened
stellar templates and the total 3rd order polynomial contribution
whose sum creates the model seen in the middle panel.}
\end{figure}

\clearpage

\begin{figure}
\epsfig{file=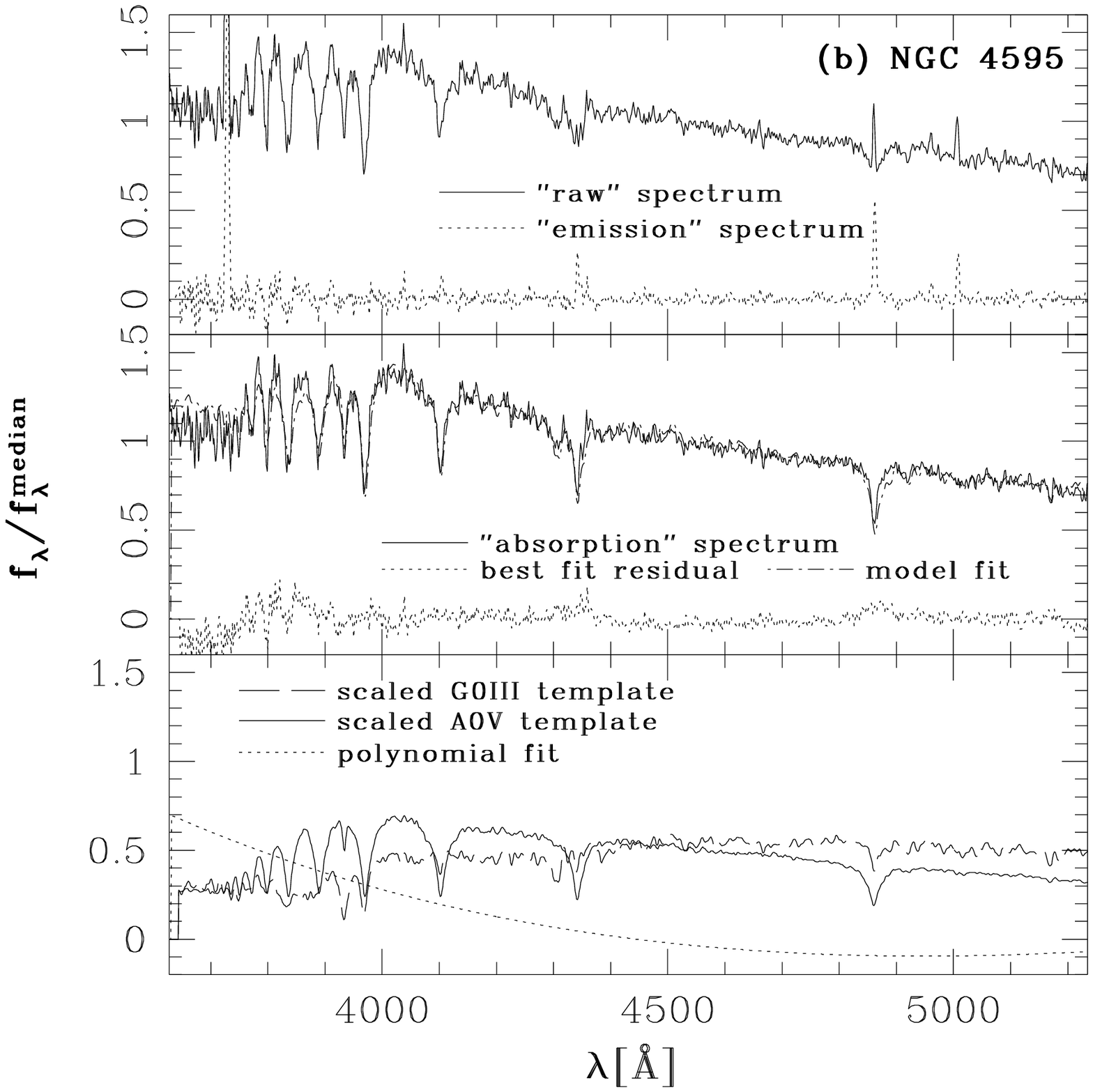,height=6.5in,width=6.5in}
\end{figure}

\clearpage

\begin{figure}
\epsfig{file=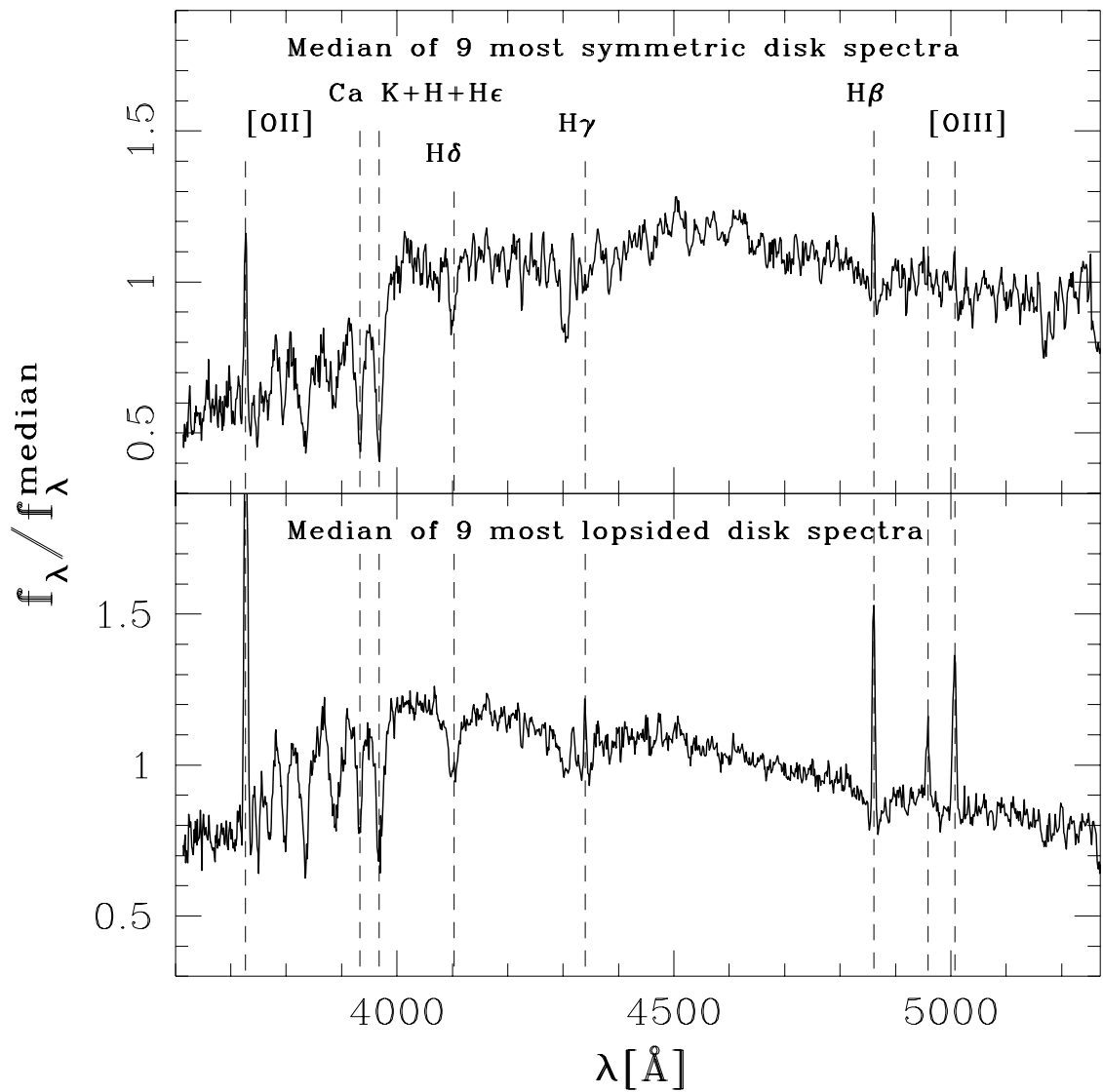,height=6.5in,width=6.5in}
\caption{ 
The composite, median spectrum of our nine most symmetric (top) and
nine most lopsided (bottom) galaxies.  Note that the lopsided
composite has a bluer continuum, stronger Balmer absorption lines,
stronger emission lines and a shallower $4000~{\rm \AA}$ break than
the symmetric composite.  The median values of $\langle \tilde{A}_1
\rangle$ for the symmetric and lopsided composites are 0.04 and 0.33
respectively.}
\end{figure}

\clearpage

\begin{figure}
\epsfig{file=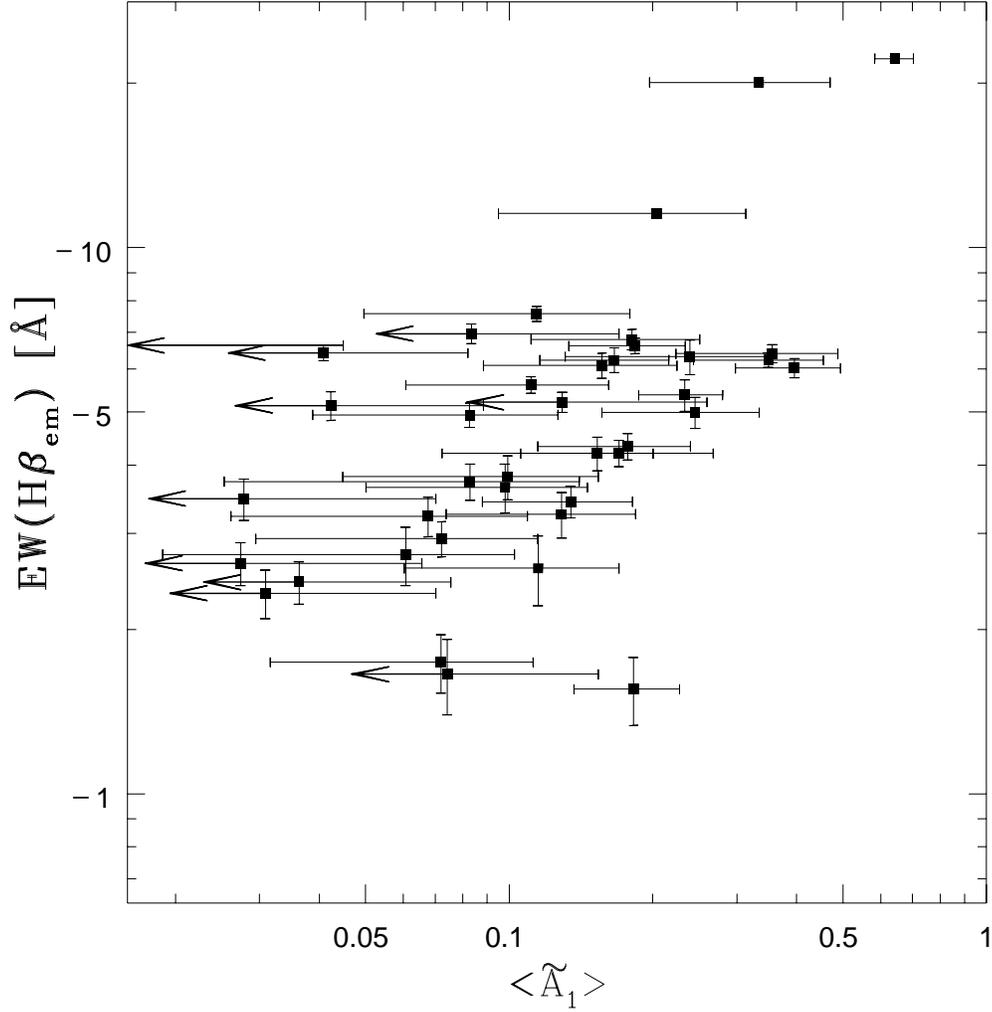,height=6.5in,width=6.5in}
\caption{
The $EW(H\beta_{em})$, integrated over the disks, as measured from the
decomposed ``emission'' spectrum of each galaxy.  Arrows on data
points indicate that the errors are larger than the values themselves
and that the values should be therefore be treated as upper
limits.}
\end{figure}

\clearpage

\begin{figure}
\epsfig{file=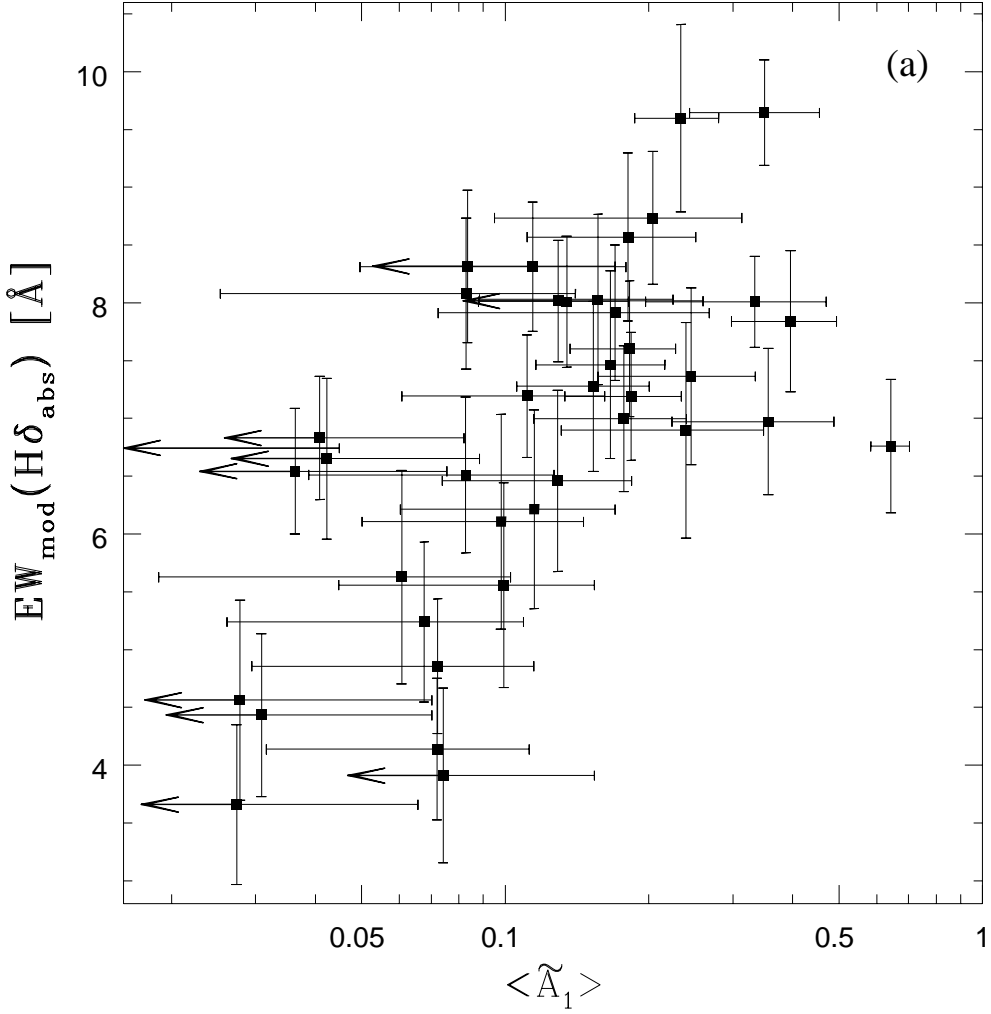,height=6.5in,width=6.5in}
\caption{
 Measures of the recent SFH vs. $\langle \tilde{A}_1\rangle$: (a)
 $EW_{mod}(H\delta_{abs})$ as measured from our best fit stellar
 template model to the ``absorption'' spectrum of the galaxy.  (b) The
 value of $D_{4000}$ as measured from our ``absorption'' spectra.  (c)
 The scaling factor of the A star template ($C_{A0V}$) in the best
 model fit to the ``absorption'' spectrum of the galaxy
 $I_{model}^{best} (\lambda)$.  Arrows on data points indicate upper
 limits.}
\end{figure}

\clearpage

\begin{figure}
\epsfig{file=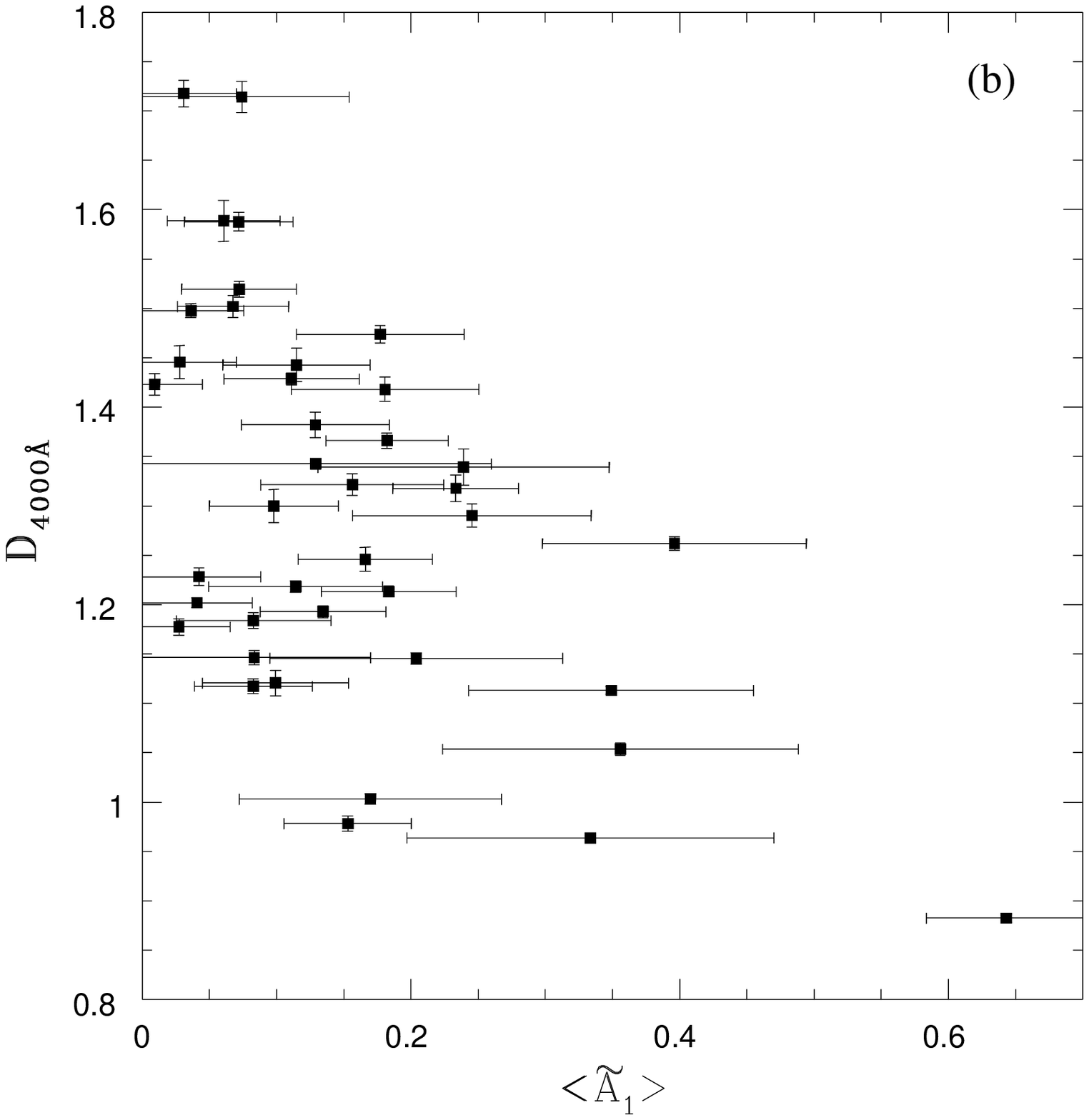,height=6.5in,width=6.5in}
\end{figure}

\clearpage

\begin{figure}

\epsfig{file=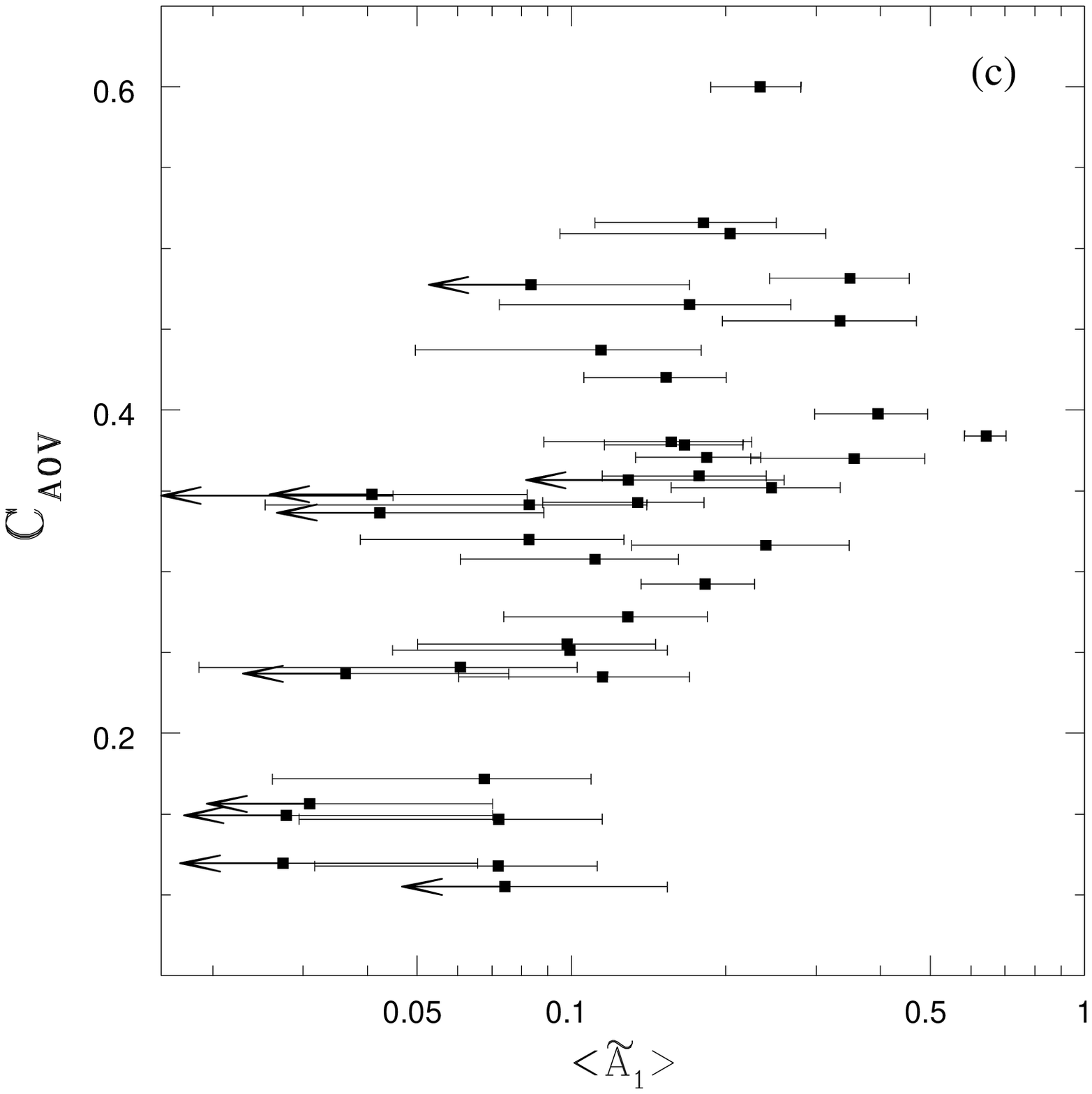,height=6.5in,width=6.5in}
\end{figure}

\clearpage

\begin{figure}
\epsfig{file=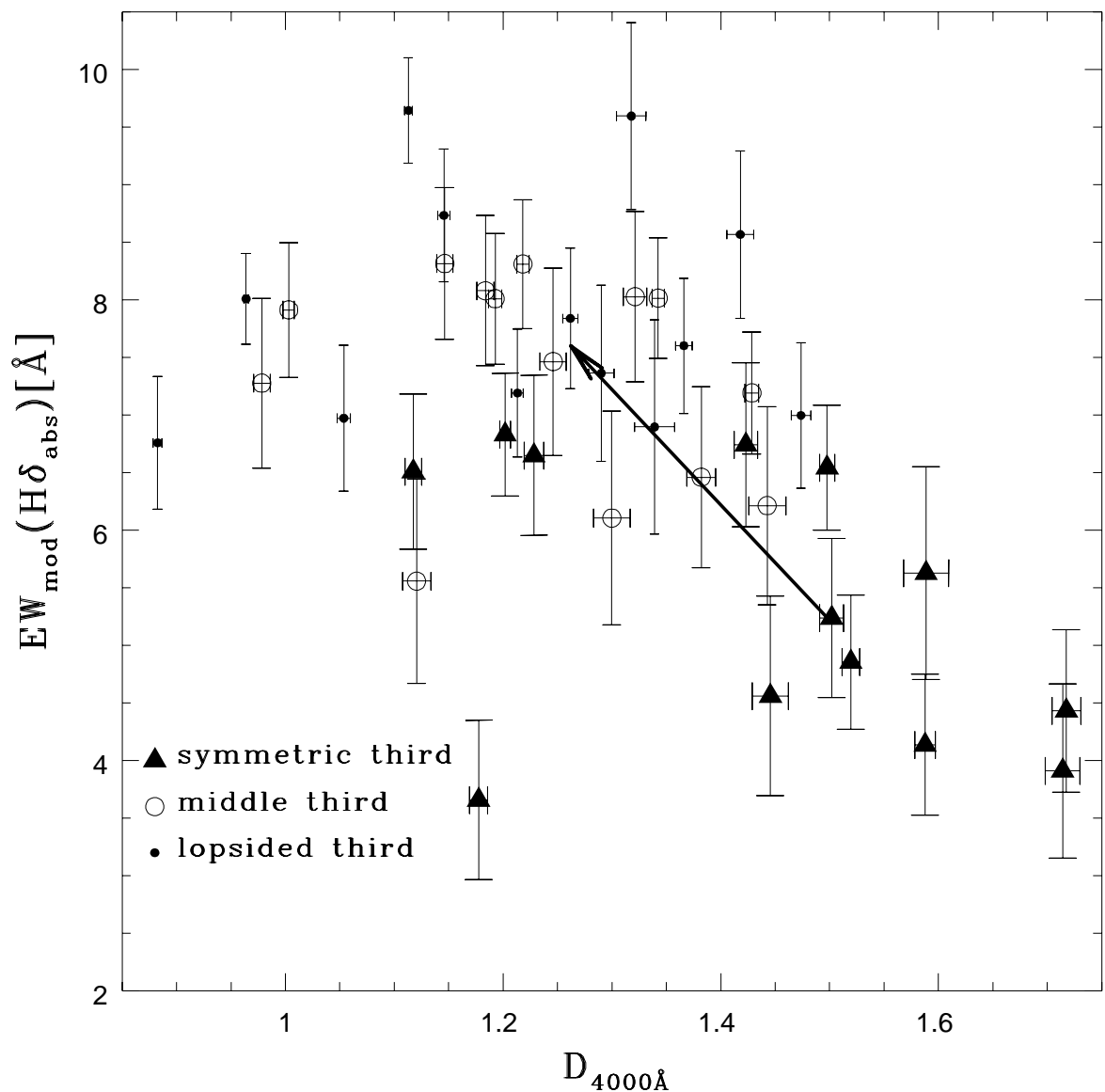,height=6.5in,width=6.5in}
\caption{ 
The values for all of the galaxies plotted in the
$EW_{mod}(H\delta_{abs})$ vs. $D_{4000}$ plane.  We split the data
into the most, middle and least lopsided thirds of the sample.  We
also overlay the boost vector connecting the median values of the
spectral indices for the most symmetric third, to the median values
for the most lopsided third.}
\end{figure}

\clearpage

\begin{figure}
\epsfig{file=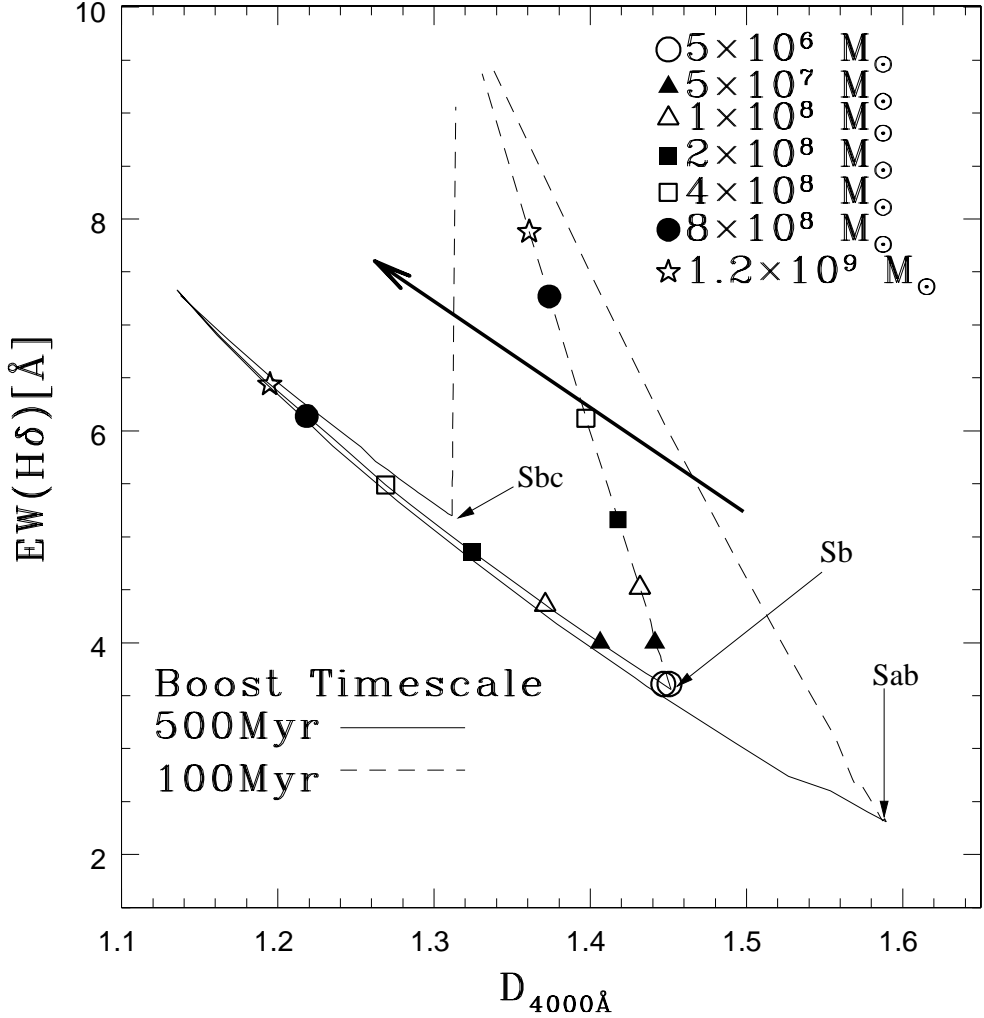,height=6.5in,width=6.5in}
\caption{ 
Model tracks for the ``underlying population $+$ boost'' EPS models.
Each track represents the evolution of $D_{4000}$ and
$EW_{mod}(H\delta_{abs})$ with boost strength for a given Hubble type
(as parameterized by $b$) and boost timescale.  The symbols represent
different masses formed in the boost in addition to that formed in the
underlying population over the past 0.5 Gyr.  We also overlay the
boost vector connecting the median values of the spectral indices for
the most symmetric third, to the median values for the most lopsided
third.  For additional reference, we plot tracks for galaxies of
Hubble type Sab, Sb, and Sbc.}
\end{figure}

\clearpage

\begin{figure}
\epsfig{file=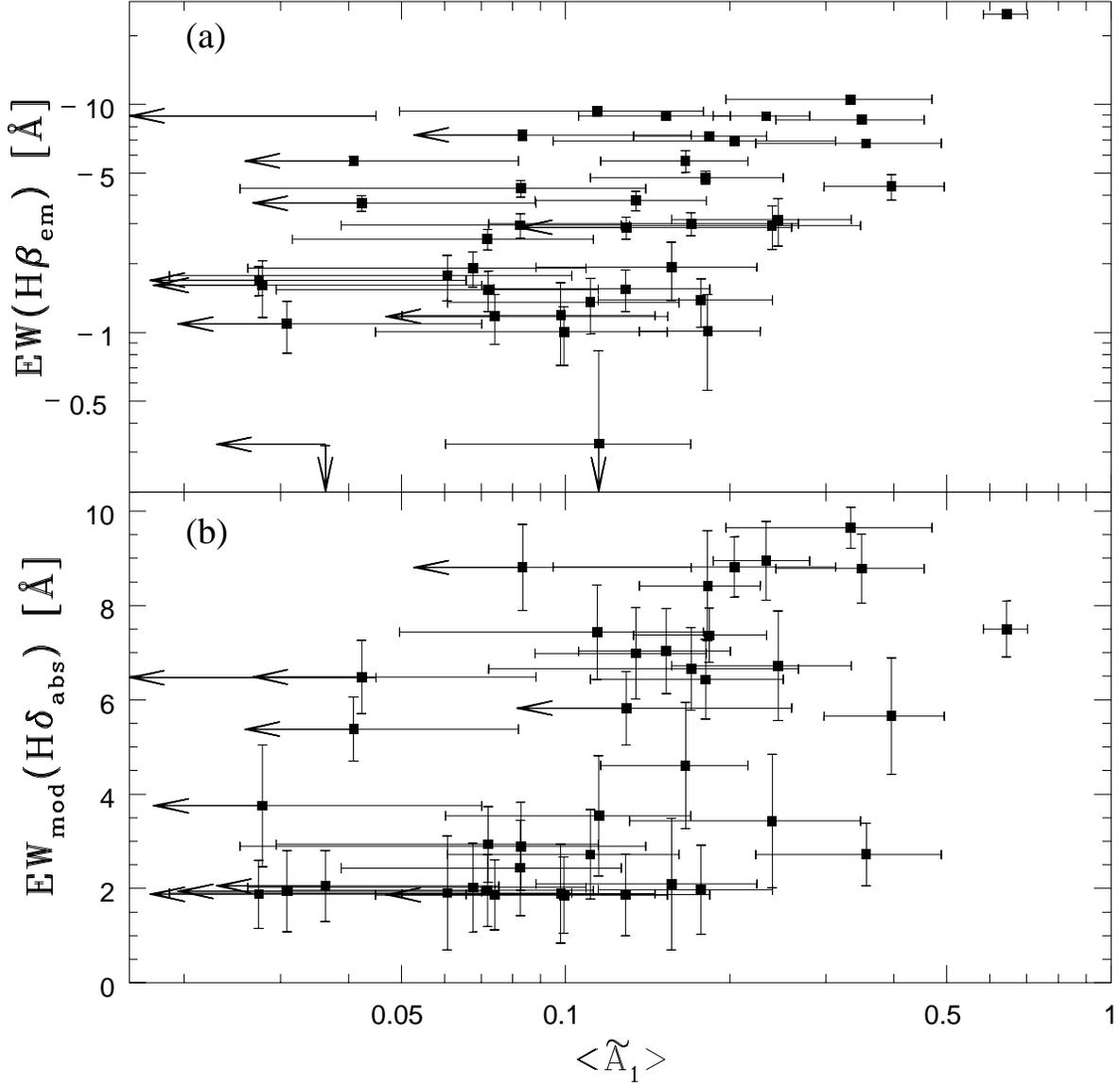,height=6.5in,width=6.5in}
\caption{ 
Measures of the current SFR and recent SFH in the nucleus vs. $\langle
\tilde{A}_1\rangle$: (a) $EW(H\beta_{em})$ as measured from the
decomposed ``emission'' spectrum of each galaxy.  (b)
$EW_{mod}(H\delta_{abs})$ as measured from our best fit stellar
template model to the ``absorption'' spectrum of the galaxy
$I_{model}^{best} (\lambda)$.  Arrows on data points indicate upper
limits.}
\end{figure}

\clearpage

\begin{deluxetable}{llllll}
\tablewidth{0pt}
\tablecaption{Sample: Summary of RC3 Parameters}
\tablehead{\colhead{{object~name}} & \colhead{{T}} &
\colhead{{D}\tablenotemark{a}} & \colhead{{v}$_{rec}$} &
\colhead{{d/D}\tablenotemark{b}} & \colhead{$L_B$}\\
\colhead{ } & \colhead{ } & \colhead{( \arcmin )} & \colhead{({\rm
$km s^{-1}$})} & \colhead{ } & \colhead{({\rm $10^{21} h^{-2} W Hz^{-1}$})
}}
\startdata
A 1219+41 & 3.0 & 1.19 & 6927 & 0.85 & 7.52$\pm$1.03 \nl
IC 0520 & 2.0 & 1.95 & 3528 & 0.79 & 6.06$\pm$0.90 \nl
IC 0900 & 4.0 & 1.62 & 7067 & 0.65 & 9.30$\pm$0.89 \nl
IC 1269 & 4.0 & 1.70 & 6116 & 0.74 & 8.21$\pm$1.57 \nl
NGC 2347 & 3.0 & 1.78 & 4422 & 0.71 & 5.32$\pm$0.79 \nl
NGC 2582 & 2.0 & 1.23 & 4439 & 1.00 & 2.79$\pm$0.35 \nl
NGC 2744 & 2.0 & 1.66 & 3428 & 0.65 & 1.62$\pm$0.33 \nl
NGC 2916 & 3.0 & 2.45 & 3730 & 0.68 & 5.50$\pm$0.76 \nl
NGC 3066 & 4.0 & 1.10 & 2049 & 0.89 & 0.78$\pm$0.12 \nl
NGC 3162 & 4.0 & 3.02 & 1298 & 0.83 & 1.41$\pm$0.18 \nl
NGC 3177 & 3.0 & 1.44 & 1302 & 0.81 & 0.46$\pm$0.06 \nl
NGC 3310 & 4.0 & 3.09 & 0980 & 0.78 & 1.84$\pm$0.18 \nl
NGC 3353 & 3.0 & 1.35 & 0944 & 0.71 & 0.22$\pm$0.03 \nl
NGC 3681 & 4.0 & 2.51 & 1239 & 0.79 & 1.49$\pm$0.48 \nl
NGC 3684 & 4.0 & 3.09 & 1163 & 0.69 & 1.57$\pm$0.50 \nl
NGC 3897 & 4.0 & 1.95 & 6411 & 1.00 & 7.93$\pm$1.17 \nl
NGC 3928 & 3.0 & 1.51 & 0982 & 1.00 & 0.24$\pm$0.03 \nl
NGC 3963 & 4.0 & 2.75 & 3186 & 0.91 & 5.23$\pm$1.67 \nl
NGC 4017 & 4.0 & 1.78 & 3454 & 0.78 & 3.72$\pm$1.18 \nl
NGC 4041 & 4.0 & 2.69 & 1234 & 0.93 & 1.27$\pm$0.16 \nl
NGC 4351 & 2.0 & 2.00 & 2310 & 0.68 & 1.79$\pm$0.26 \nl
NGC 4412 & 3.0 & 1.41 & 2294 & 0.89 & 1.43$\pm$0.21 \nl
NGC 4430 & 3.0 & 2.29 & 1443 & 0.89 & 0.85$\pm$0.13 \nl
NGC 4595 & 3.0 & 1.74 & 0633 & 0.65 & 0.12$\pm$0.18 \nl
NGC 4639 & 4.0 & 2.75 & 1010 & 0.68 & 0.59$\pm$0.05 \nl
NGC 4814 & 3.0 & 3.09 & 2513 & 0.74 & 2.48$\pm$0.50 \nl
NGC 4911 & 4.0 & 1.45 & 7970 & 0.91 & 12.04$\pm$1.53 \nl
NGC 5218 & 3.0 & 1.82 & 2807 & 0.69 & 2.31$\pm$1.02 \nl
NGC 5614 & 2.0 & 2.45 & 3892 & 0.83 & 7.50$\pm$0.95 \nl
NGC 5653 & 3.0 & 1.74 & 3567 & 0.74 & 4.63$\pm$0.59 \nl
NGC 5713 & 4.0 & 2.75 & 1883 & 0.89 & 3.32$\pm$0.49 \nl
NGC 5915 & 4.0 & 1.82 & 5570 & 1.00 & 2.04$\pm$0.26 \nl
NGC 5923 & 2.0 & 1.74 & 2291 & 0.72 & 5.20$\pm$0.88 \nl
NGC 5936 & 3.0 & 1.45 & 4004 & 0.89 & 4.76$\pm$0.60 \nl
NGC 6012 & 2.0 & 2.09 & 1854 & 0.72 & 1.69$\pm$0.21 \nl
NGC 6195 & 3.0 & 1.55 & 9000 & 0.69 & 12.89$\pm$2.19 \nl
NGC 6574 & 4.0 & 1.41 & 2282 & 0.78 & 2.02$\pm$0.26 \nl
NGC 6711 & 4.0 & 1.35 & 4671 & 0.93 & 3.63$\pm$0.50 \nl
NGC 6824 & 3.0 & 1.70 & 3337 & 0.69 & 3.58$\pm$0.72 \nl
NGC 7177 & 3.0 & 3.10 & 1150 & 0.64 & 0.99$\pm$0.12 \nl
\enddata
\tablenotetext{a}{the diameter of the galaxy out to the isophote of surface
brightness 25 mag/asec$^2$; $D=0.1\arcmin \times 10^{D_{25}}$}
\tablenotetext{b}{the ratio of the minor to major axis as defined in the
RC3 catalog; $d/D=10^{-R_{25}}$}
\end{deluxetable}

\clearpage

\begin{deluxetable}{llllllll}
\tablewidth{0pt}
\tablecaption{Analysis~Results}
\tablehead{\colhead{{object~name}} & \colhead{${\rm R}_{exp}$} & \colhead{${\rm R}_{max}/{\rm R}_{exp}$} & \colhead{$<${\~{A}}$_1>$} & \colhead{$EW(H\beta_{em})$} & \colhead{$EW_{mod}(H\delta_{abs})$} & \colhead{$C_{A0V}$} & \colhead{$D_{4000}$
}\\
\colhead{ } & \colhead{(\arcsec)} & \colhead{ } & \colhead{ } & \colhead{($~{\rm \AA}$)} & \colhead{($~{\rm \AA}$)} & \colhead{ } & \colhead{ 
}}
\startdata
A 1219+41 & 19.7 & 3.15 & 0.083$\pm$0.057 & -3.7$\pm$0.3 & 8.1$\pm$0.6 & 0.34 & 1.18$\pm$0.01 \nl
IC 0520 & 19.8 & 3.82 & 0.031$\pm$0.039 & -2.3$\pm$0.2 & 4.4$\pm$0.7 & 0.16 & 1.72$\pm$0.01 \nl
IC 0900 & 11.6 & 6.54 & 0.156$\pm$0.068 & -6.1$\pm$0.3 & 8.0$\pm$0.7 & 0.38 & 1.32$\pm$0.01 \nl
IC 1269 & 13.0 & 2.93 & 0.239$\pm$0.108 & -6.3$\pm$0.5 & 6.9$\pm$0.9 & 0.32 & 1.34$\pm$0.02 \nl
NGC 2347 & 10.6 & 5.34 & 0.083$\pm$0.044 & -4.9$\pm$0.3 & 6.5$\pm$0.7 & 0.32 & 1.12$\pm$0.01 \nl
NGC 2582 & 20.0 & 2.82 & 0.061$\pm$0.042 & -2.7$\pm$0.3 & 5.6$\pm$0.9 & 0.24 & 1.59$\pm$0.02 \nl
NGC 2744 & 10.8 & 4.28 & 0.233$\pm$0.047 & -5.4$\pm$0.4 & 9.6$\pm$0.8 & 0.60 & 1.32$\pm$0.01 \nl
NGC 2916 & 23.0 & 3.29 & 0.177$\pm$0.062 & -4.3$\pm$0.2 & 7.0$\pm$0.6 & 0.36 & 1.47$\pm$0.01 \nl
NGC 3066 & 13.7 & 3.40 & 0.356$\pm$0.132 & -6.4$\pm$0.2 & 7.0$\pm$0.6 & 0.37 & 1.05$\pm$0.01 \nl
NGC 3162 & 26.0 & 2.91 & 0.181$\pm$0.070 & -6.8$\pm$0.3 & 8.6$\pm$0.7 & 0.52 & 1.42$\pm$0.01 \nl
NGC 3177 & 8.6 & 6.58 & 0.183$\pm$0.050 & -6.6$\pm$0.2 & 7.2$\pm$0.5 & 0.37 & 1.213$\pm$0.005 \nl
NGC 3310 & 9.92 & 7.63 & 0.334$\pm$0.137 & -20.1$\pm$0.2 & 8.0$\pm$0.4 & 0.45 & 0.963$\pm$0.002 \nl
NGC 3353 & 12.1 & 4.23 & 0.643$\pm$0.059 & -22.2$\pm$0.2 & 6.8$\pm$0.6 & 0.38 & 0.883$\pm$0.004 \nl
NGC 3681 & 24.9 & 3.03 & 0.027$\pm$0.038 & -2.6$\pm$0.2 & 3.7$\pm$0.7 & 0.12 & 1.18$\pm$0.01 \nl
NGC 3684 & 16.8 & 4.51 & 0.042$\pm$0.046 & -5.1$\pm$0.3 & 6.6$\pm$0.7 & 0.34 & 1.23$\pm$0.01 \nl
NGC 3897 & 12.6 & 4.49 & 0.099$\pm$0.054 & -3.8$\pm$0.3 & 5.6$\pm$0.9 & 0.25 & 1.12$\pm$0.01 \nl
NGC 3928 & 18.7 & 4.03 & 0.009$\pm$0.036 & -6.6$\pm$0.3 & 6.7$\pm$0.7 & 0.35 & 1.42$\pm$0.01 \nl
NGC 3963 & 25.4 & 2.98 & 0.135$\pm$0.047 & -3.4$\pm$0.2 & 8.0$\pm$0.6 & 0.34 & 1.19$\pm$0.01 \nl
NGC 4017 & 16.0 & 3.89 & 0.083$\pm$0.087 & -7.0$\pm$0.3 & 8.3$\pm$0.7 & 0.48 & 1.15$\pm$0.01 \nl
NGC 4041 & 10.8 & 6.97 & 0.041$\pm$0.041 & -6.4$\pm$0.2 & 6.8$\pm$0.5 & 0.35 & 1.20$\pm$0.005 \nl
NGC 4351 & 18.1 & 4.18 & 0.153$\pm$0.047 & -4.2$\pm$0.3 & 7.3$\pm$0.7 & 0.42 & 0.98$\pm$0.01 \nl
NGC 4412 & 23.7 & 2.90 & 0.396$\pm$0.098 & -6.0$\pm$0.2 & 7.8$\pm$0.6 & 0.40 & 1.26$\pm$0.01 \nl
NGC 4430 & 28.3 & 2.67 & 0.245$\pm$0.089 & -5.1$\pm$0.3 & 7.4$\pm$0.8 & 0.35 & 1.29$\pm$0.01 \nl
NGC 4595 & 16.3 & 3.82 & 0.170$\pm$0.097 & -4.2$\pm$0.2 & 7.9$\pm$0.6 & 0.46 & 1.003$\pm$0.005 \nl
NGC 4639 & 23.4 & 3.24 & 0.072$\pm$0.043 & -2.9$\pm$0.2 & 4.9$\pm$0.6 & 0.15 & 1.52$\pm$0.01 \nl
NGC 4814 & 13.5 & 5.58 & 0.068$\pm$0.041 & -3.2$\pm$0.3 & 5.2$\pm$0.7 & 0.17 & 1.50$\pm$0.01 \nl
NGC 4911 & 14.5 & 3.54 & 0.129$\pm$0.055 & -3.2$\pm$0.3 & 6.5$\pm$0.8 & 0.27 & 1.38$\pm$0.01 \nl
NGC 5218 & 8.7 & 4.86 & 0.182$\pm$0.046 & -1.6$\pm$0.2 & 7.6$\pm$0.6 & 0.29 & 1.37$\pm$0.01 \nl
NGC 5614 & 11.9 & 4.76 & 0.074$\pm$0.080 & -1.7$\pm$0.3 & 3.9$\pm$0.8 & 0.10 & 1.71$\pm$0.01 \nl
NGC 5653 & 6.8 & 5.63 & 0.129$\pm$0.131 & -5.2$\pm$0.2 & 8.0$\pm$0.5 & 0.36 & 1.343$\pm$0.005 \nl
NGC 5713 & 14.6 & 4.27 & 0.349$\pm$0.106 & -6.2$\pm$0.2 & 9.6$\pm$0.5 & 0.48 & 1.113$\pm$0.003 \nl
NGC 5915 & 5.0 & 3.48 & 0.204$\pm$0.109 & -11.5$\pm$0.2 & 8.7$\pm$0.6 & 0.51 & 1.146$\pm$0.005 \nl
NGC 5923 & 15.2 & 2.52 & 0.115$\pm$0.055 & -2.6$\pm$0.4 & 6.2$\pm$0.9 & 0.23 & 1.44$\pm$0.02 \nl
NGC 5936 & 11.7 & 3.61 & 0.114$\pm$0.065 & -7.6$\pm$0.2 & 8.3$\pm$0.6 & 0.44 & 1.218$\pm$0.005 \nl
NGC 6012 & 12.2 & 3.46 & 0.028$\pm$0.042 & -3.5$\pm$0.3 & 4.6$\pm$0.9 & 0.15 & 1.45$\pm$0.02 \nl
NGC 6195 & 11.2 & 4.17 & 0.098$\pm$0.048 & -3.6$\pm$0.4 & 6.1$\pm$0.9 & 0.25 & 1.30$\pm$0.02 \nl
NGC 6574 & 8.6 & 4.46 & 0.111$\pm$0.050 & -5.6$\pm$0.2 & 7.2$\pm$0.5 & 0.31 & 1.43$\pm$0.01 \nl
NGC 6711 & 9.2 & 3.13 & 0.166$\pm$0.050 & -6.2$\pm$0.3 & 7.5$\pm$0.8 & 0.38 & 1.25$\pm$0.01 \nl
NGC 6824\tablenotemark{a} & 14.5 & 4.64 & 0.068$\pm$0.1 & -2.4$\pm$0.2 & 6.5$\pm$0.5 & 0.24 & 1.50$\pm$0.01 \nl
NGC 7177\tablenotemark{a} & 10.9 & 5.66 & 0.140$\pm$0.1 & -1.7$\pm$0.2 & 4.1$\pm$0.6 & 0.12 & 1.59$\pm$0.01 \nl
\label{tbl:table2.7.23.99.in}
\enddata
\tablenotetext{a}{The fourth column is the averages of two images}
\end{deluxetable}

\end{document}